\documentclass[12pt]{article}
\usepackage{amssymb, amsmath, amsfonts}

\newtheorem{thm}{Theorem}[section]

\newtheorem{lem}[thm]{Lemma}
\newtheorem{rem}[thm]{Remark}

\newtheorem{ex}{Example}

\begin{document}

\title{RECURSION OPERATOR FOR A SYSTEM ADMITTING LAX REPRESENTATION WITH NON-RATIONAL LAX FUNCTION}
\author{ K. Zheltukhin\\
{\small Department of Mathematics, Faculty of Sciences}\\
{\small Middle East Technical University, 06800 Ankara-Turkey}\\
{\it email:zheltukh@metu.edu.tr}}

\date{\vspace{-5ex}}

\begin{titlepage}
\maketitle

\begin{abstract}
A recursion operator  is constructed for a hydrodynamic type system admitting dispersionless Lax representation with non-rational Lax function.
\end{abstract}

\end{titlepage}

\section{Introduction}
In the present paper we construct a recursion operator for a hydrodynamic type system admitting Lax representation with a non-rational Lax function.
Let us note that in the case of rational and polynomial Lax functions we can construct a recursion operator  following \cite{GKS}. 
 We write a recursion relation 
 \begin{equation}
 L_{t_{n+1}}=LL_{t_n}+\{R_n,L\}
 \end{equation}
  between the symmetries. In case of a polynomial or rational  Lax function the form  of the remainder $R_n$ can be predicted. 
  So, we can use the above recursion relation to find a recursion operator, see  \cite{GKS}-\cite{Zhel} for detail.
 For the non-rational Lax function  it is not possible  to predict the form of the remainder $R_n$ and apply the method of \cite{GKS}.
 So we construct the recursion operator analysing the Lax equation itself.  For constructions of recursion operators of some other classes of hydrodynamic 
 type systems see also \cite{fg}-\cite{tes}. 
 
Let us give necessary definitions. We introduce the algebra of Laurent series 
\begin{equation}
{\cal A}=\left\{ \sum_{-\infty}^\infty u_ip^i: u_i \quad\mbox{-- smooth rapidly decreasing at infinity functions} \right\},
\end{equation}
with the Poisson bracket given by 
\begin{equation}
\{f,g\}=\frac{\partial f}{\partial p}\frac{\partial g}{\partial x}-\frac{\partial f}{\partial x}\frac{\partial g}{\partial p}.
\end{equation}
Using the Gelfond-Dikkii construction  \cite{GelDik} we can write the hierarchy of integrable equations on the algebra $\cal A$.
Equations of such type have applications in the topological field theories \cite{D} and appear as equations describing slow varying quasi periodic solutions of non-linear integrable equations, see \cite{CEM}-\cite{T3}.
We consider non-rational Lax function  
\begin{equation}\label{lax_f}
L=\mu -m\ln(\mu-c^1)+\ln(\mu -c^2)+\dots +\ln(\mu-c^{m+1}).
\end{equation}
The corresponding Lax equation 
\begin{equation}
L_t=\{(L^2)_{\ge 0},L\}
\end{equation}
leads to the system 
\begin{equation}\label{waterbag1}
c^j_t=\partial_x\left( \frac{(c^j)^2}{2} +mc^1- c^2-\dots -c^{m+1}\right), 
\end{equation}
where $j=1,2,\dots (m+1)$, see \cite{sb},\cite{Pav} and references there in.
We have a hierarchy of symmetries for the above equation given by 
\begin{equation}
L_t=\{(L^n)_{\ge 0},L\} \qquad n=1,2,\dots
\end{equation}
As we show the above hierarchy admits the following recursion operator 
\begin{equation}\label{rec_operator_waterbag1}
{\cal R}=A\partial_x^{-1},
\end{equation}
where matrix $A=(\gamma_{ij})$ has entries
$$
\gamma_{11}=c^1_x+ \sum_{j=2}^{m+1} \frac{c^1_x-c^j_x}{c^1-c^j},\quad
\gamma_{1k}=-\frac{c^1_x-c^k_x}{c^1-c^k},\quad
\gamma_{k1}=m\frac{c^1_x-c^k_x}{c^1-c^k},
$$
$$
\gamma_{kk}=c^k_x-m\frac{c^1_x-c^k_x}{c^1-c^k} +\sum_{j=2,j\ne k}^{m+1} \frac{c^k_x-c^j_x}{c^k-c^j}, \qquad 
\gamma_{ki}= -\frac{c^k_x-c^i_x}{c^k-c^i}
$$
$ k\ne i,\,\, \mbox{and}\,\, k,i=2,3,\dots, m+1.$

The paper is organized as follows. In  Section 2 we give general construction of the recursion operator and Section 3 we consider several examples.

\section{Evaluation of recursion operator}

Let us introduce  new variables 
\begin{equation}
c^1=u \quad \mbox{and} \quad v^{j-1}=c^1-c^j \quad j=2,3,\dots (m+1) \, .
\end{equation}
In new variables the system (\ref{waterbag1}) takes the form
\begin{equation}\label{waterbag2}
\begin{array}{l}
u_t=uu_x+v^1_x+\dots v^n_x\\
v^1_t=v^1u_x+(u-v^1)v^1_x\\
\dots\\
v^m_t=v^mu_x+(u-v^m)v^m_x\\
\end{array}
\end{equation}
The system (\ref{waterbag2}) admits a Lax representation 
\begin{equation}\label{lax_eqn_waterbag2}
L_t=\{(L^2)_{\ge 1},L\}
\end{equation}
with Lax function 
\begin{equation}
L=p+u + \ln \left (1+\frac{v^1}{p}\right)+\ln \left(1+\frac{v^2}{p}\right)+\dots +\ln \left(1+\frac{v^m}{p}\right).
\end{equation}
Thus we have the whole  hierarchy of symmetries for the system (\ref{waterbag2}) given by 
\begin{equation}\label{Lax_n}
L_{t_n}=\{(L^n)_{\ge 1},L\} \quad n=1,2,\dots
\end{equation}

Let us construct a recursion operator for the above hierarchy of symmetries.
We construct the recursion operator by direct analysis of the Lax representation.
Let 
\begin{equation}\label{L^n}
L^n=a_np^n+a_{n-1}p^{n-1}+\dots a_1p+a_0+a_{-1}p^{-1}+\dots
\end{equation} 
The next two lemmas give some relations between coefficients of $L^n$ and 
$$
L_{t_n}=u_{t_n}+\frac{ v^1_{t_n}}{p+1}+\dots +   \frac{ v^m_{t_n}}{p+1}.
$$.

\begin{lem}\label{lem_sum}
For any $k=2,3 \dots m$ and any $n=2,3,\dots$ the following 
equality holds 
\begin{equation}
\sum_{i=1}^n (-1)^{(i-1)} a_i(v^k)^i=\partial^{-1}_x  v^k_{t_n} 
\end{equation} 
\end{lem}
{\bf Proof.}
Using (\ref{L^n}) we can write the equation (\ref{Lax_n}) as
$$
\begin{array}{l}
u_{t_n} +\frac{v^1_{t_n}}{p+v^1}+\dots+\frac{v^m_{t_n}}{p+v^m}= \\
\hspace{0.5cm} (na_np^{n-1}+\dots+2a_2p +a_1)\left(u_{x} +\frac{v^1_{x}}{p+v^1}+\dots+\frac{v^m_{x}}{p+v^m}\right)-\\
\hspace{0.5cm} (a_{n,x}p^{n}+\dots+a_{2,x}p^2 +a_{1,x})\left(1-\frac{v^1}{p(p+v^1)}-\dots-\frac{v^m}{p(p+v^m)}\right)\\
\end{array}
$$
Multiplying the above equation by $(p+v_1)(p+v_2)\dots(p+v_m)$ and then substituting $p=-v_k$ we obtain
$$
v^k_{t_n}=\sum_{i=1}^n (-1)^{i-1}i a_i (v^k)^{i-1}v^k_x+\sum_{i=1}^n(-1)^{i-1}a_{i,x}(v^k)^i.
$$
That is 
$$
v^k_{t_n}=\left(\sum_{i=1}^n (-1)^{i-1} a_i(v^k)^i\right)_x. \quad \Box
$$

\begin{lem}\label{lem_a0}
For  any $n=2,3,\dots$ the following 
equality holds $a_0=\partial^{-1}_x u_{t_n} . $
\end{lem}
{\bf Proof.}
The Lax equation (\ref{Lax_n}) can be written as 
$$
L_{t_n}=\{(L^n)_{\le 0},L\} \quad n=1,2,\dots
$$
Using (\ref{L^n}) and collecting coefficients of zero power of $p$ in the above equations 
we have $u_{t_n}=a_{0,x}$. $\Box$

The above lemmas allow us  to express the coefficients of $(L^{(n+1)}_{>0})_p$ and $(L^{(n+1)}_{>0})_x$ in terms of coefficients of $L^n_{\ge 0}$ and  
$L_{t_n}$.

\begin{lem}\label{coefficients}
Let 
\begin{equation}
\frac{1}{n+1}\left( L^{(n+1)}_{\ge 1}\right)_p=b_np^{n-1}+\dots +b_2p+b_1. 
\end{equation}
Then 
\begin{equation}\label{b_r}
b_r=a_{r-1}+\sum_{k=1}^{m} \sum_{j=0}^{r-1} (v^k)^{-j}a_{r-j} +  \sum_{k=1}^{m} (v^k)^{-r} \partial^{-1}_x v^k,
\end{equation}
where $r=1,2,\dots m$.

Let 
\begin{equation}
\frac{1}{n+1}\left(L^{(n+1)}_{\ge 1}\right)_x=d_np^n+\dots +d_2p^2+d_1p. 
\end{equation}
Then 
\begin{equation}\label{d_r}
d_r=u_xa_r+\sum_{k=1}^{m} \sum_{j=0}^{r-1} (v^k)^{-j-1}v^k_xa_{r-j} 
+ \sum_{k=1}^{m} (v^k)^{-r-1}v^k_x\partial^{-1}_x v^k,
\end{equation}
where $r=1,2,\dots m$.
\end{lem}
{\bf Proof.}
We have 
$$
\frac{1}{n+1}\left(L^{(n+1)}_{\ge 1}\right)_p=\left(L^{(n)}_{\ge 0} L_p\right)_{\ge 0}.
$$
That is 
$$
\frac{1}{n+1}\left (L^{(n+1)}_{\ge 1}\right )_p=\left((a_np^n+\dots+a_0)\left(u_x+\sum_{k=1}^m \frac{v^k_x}{p+v_k}  \right)\right)_{\ge 0}.
$$
For each $k=1,\dots m$, we expand  $\displaystyle{\frac{1}{p+v^k}}$ as series in terms of $p^{-1}$ around $p=\infty$ and multiply with
$(a_np^n+\dots+a_0)$. Collecting coefficients of $p^k$, $k=1,\dots m$, in the above equality and using Lemma~\ref{lem_sum} we obtain
 formula (\ref{b_r}). The formula (\ref{d_r}) is obtained in the same way.  $\Box$
 
 Using the above lemmas we find a recursion operator for the hierarchy (\ref{Lax_n}).
 
\begin{lem}\label{rec_operator}
The recursion operator for the system (\ref{waterbag2}) can be written as  ${\cal R}=C\partial_x^{-1},$ where $C$ is an $(m+1)\times(m+1)$ matrix. It is convenient to write the matrix $C$ as a sum of two matrices,  $C= (A+B)$.
 The matrix $A=(\alpha_{ij})$ has entries 
$$
\alpha_{11}=u_x;
$$
$$
 \alpha_{1(j+1)}=v^j_x(v^j)^{-1}, \,j=1,2,\dots, m;
$$
$$
\quad \alpha_{(j+1)1}=v^j_x, \,j=1,2,\dots, m;
$$
$$
\alpha_{(j+1)(j+1)}=(u_x-v^j_x), \,j=1,2,\dots, m; 
$$
$$
\alpha_{(i+1)(j+1)}=0  \quad  i\ne j  \quad i,j=1,2,\dots, m;
$$
The matrix $B=(\beta_{ij})$ has entries 
$$
\beta_{11}=0;
$$
$$
 \beta_{1(j+1)}=0, \,j=1,2,\dots, m;
$$
$$
 \beta_{(j+1)1}=0, \,j=1,2,\dots, m;
$$
$$
\beta_{(j+1)(j+1)}=\sum_{k=1,k\ne j}^m \frac{v^k_x-v^k(v^j)_x(v^j)^{-1}}{v^k-v^j} \quad j=1,2,\dots, m; 
$$
$$
\beta_{(i+1)(j+1)}=\frac{v^i_x-v^iv^j_x(v^j)^{-1}}{v^j-v^i} \quad  i\ne j  \quad i,j=1,2,\dots, m;
$$

\end{lem}
{\bf Proof.} 
Using notations of Lemma \ref{coefficients} the Lax equation (\ref{Lax_n}) can be written as 
\begin{equation}
\begin{array}{ll}
u_{t_{n+1}} &\displaystyle{ + \sum_{k=1}^m \frac{v^k_{t_{n+1}}}{p+v^k}= }\\
 & \displaystyle{ (n+1)(b_np^{n-1}+\dots +b_2p+b_1)\left(u_x+\sum_{k=1}^m \frac{v^k_x}{p+v^k}\right)-}\\
 & \displaystyle{ (n+1)(d_np^{n}+\dots +d_2p^2+d_1)\left(1-\sum_{k=1}^m \frac{v^k}{p(p+v^k)}\right)}.\\
\end{array}
\end{equation}
 We multiply the above equation by $(p+v^1)(p+v^2)\dots(p+v^m)$ and substitute expressions for $b_i,d_i$, $i=1,2,\dots n$, given in Lemma \ref{coefficients}. Equating coefficients of $p^k$, $k=1,2,\dots m,$ we obtain a  system of equations  linear with respect to $v^k_{t_{n+1}}$, $k=1,2,\dots m$.
Solving the system we obtain the recursion operator given above. $\Box$  
 
\begin{rem}
Let us define  vector  $V=(u, v^1, v^2, \dots, v^m)$ and write the  system (\ref{waterbag2}) as 
\begin{equation}
V_t=K(V,V_x).
\end{equation} 
It follows by direct calculations that the constructed above operator satisfies the criteria for recursion operators 
\begin{equation}
{\cal R}_t={\mathbb D}_K{\cal R}-{\cal R} {\mathbb D}_K, 
\end{equation} 
where $\mathbb D_K$ is the Freshet derivative of $K$.
\end{rem} 

Returning to the original variables $c^1,\dots c^{m+1}$ we obtain the recursion operator (\ref{rec_operator_waterbag1}).   

\section{Examples}
 
Let us consider some examples. We give examples in variables $c^1,c^2,\dots, c^{m+1}$. 
 
\begin{ex}
Let us consider equation (\ref{waterbag1}) with $m=1$.  The equation becomes
\begin{equation}
\begin{array}{l}
c^1_t=c^1c^1_x+c^1_x-c^2_x\\
c^2_t=c^2c^2_x+c^1_x-c^2_x\\
\end{array}
\end{equation}
The above system admits the recursion operator 
\begin{equation}
\left(
\begin{array}{ll}
 c^1_x+\frac{c^1_x-c^2_x}{c^1-c^2}&-\frac{c^1_x-c^2_x}{c^1-c^2} \\
\frac{c^1_x-c^2_x}{c^1-c^2}& c^2_x -\frac{c^1_x-c^2_x}{c^1-c^2}\\
\end{array}
\right)\partial_x^{-1}.
\end{equation}
\end{ex} 

\begin{ex}
Let us consider equation (\ref{waterbag1}) with $m=2$.  The equation becomes
\begin{equation}
\begin{array}{l}
c^1_t=c^1c^1_x+2c^1_x-c^2_x-c^3_x\\
c^2_t=c^2c^2_x+2c^1_x-c^2_x-c^3_x\\
c^3_t=c^3c^3_x+2c^1_x-c^2_x-c^3_x\\
\end{array}
\end{equation}
The above system admits the recursion operator 
\begin{equation}
\left(
\begin{array}{lll}
 c^1_x+\frac{c^1_x-c^2_x}{c^1-c^2}+\frac{c^1_x-c^3_x}{c^1-c^3}&-\frac{c^1_x-c^2_x}{c^1-c^2}& -\frac{c^1_x-c^3_x}{c^1-c^3}\\
2\frac{c^1_x-c^2_x}{c^1-c^2}& c^2_x -2\frac{c^1_x-c^2_x}{c^1-c^2}+\frac{c^2_x-c^3_x}{c^2-c^3}&-\frac{c^2_x-c^3_x}{c^2-c^3}\\
2\frac{c^1_x-c^3_x}{c^1-c^3}& -\frac{c^3_x-c^2_x}{c^3-c^2}&  c^3_x -2\frac{c^1_x-c^3_x}{c^1-c^3}+\frac{c^3_x-c^2_x}{c^3-c^2} \\
\end{array}
\right)\partial_x^{-1}.
\end{equation}
\end{ex} 

\begin{ex}
Let us consider equation (\ref{waterbag1}) with $m=3$.  The equation becomes
\begin{equation}
\begin{array}{l}
c^1_t=c^1c^1_x+3c^1_x-c^2_x-c^3_x-c^4_x\\
c^2_t=c^2c^2_x+3c^1_x-c^2_x-c^3_x-c^4_x\\
c^3_t=c^3c^3_x+3c^1_x-c^2_x-c^3_x-c^4_x\\
c^4_t=c^4c^4_x+3c^1_x-c^2_x-c^3_x-c^4_x\\
\end{array}
\end{equation}
The above system admits the recursion operator 

\begin{equation}
\left(
\begin{array}{llll}
 c^1_x &-\frac{c^1_x-c^2_x}{c^1-c^2}& -\frac{c^1_x-c^3_x}{c^1-c^3}& -\frac{c^1_x-c^4_x}{c^1-c^4}\\
3\frac{c^1_x-c^2_x}{c^1-c^2}& c^2_x -3\frac{c^1_x-c^2_x}{c^1-c^2}&-\frac{c^2_x-c^3_x}{c^2-c^3}&-\frac{c^2_x-c^4_x}{c^2-c^4}\\
3\frac{c^1_x-c^3_x}{c^1-c^3}& -\frac{c^3_x-c^2_x}{c^3-c^2}&  c^3_x -3\frac{c^1_x-c^3_x}{c^1-c^3}& -\frac{c^3_x-c^4_x}{c^3-c^4}\\
3\frac{c^1_x-c^4_x}{c^1-c^4}& -\frac{c^4_x-c^2_x}{c^4-c^2}& -\frac{c^4_x-c^3_x}{c^4-c^3}& c^4_x-3\frac{c^4_x-c^1_x}{c^4-c^1}\\
\end{array}
\right)\partial_x^{-1}+.
\end{equation}
$$
\left(
\begin{array}{llll}
\sum\limits_{j=2}^4\frac{c^1_x-c^j_x}{c^1-c^j}&0&0&0\\
0& \sum\limits_{j=2,j\ne 2}^4\frac{c^2_x-c^j_x}{c^2-c^j}&0&0\\
0& 0& \sum\limits_{j=2,j\ne 3}^4\frac{c^3_x-c^j_x}{c^3-c^j}& 0\\
0&0&0&  \sum\limits_{j=2,j\ne 4}^4\frac{c^4_x-c^j_x}{c^4-c^j}\\
\end{array}
\right)\partial_x^{-1}
$$
\end{ex}

\end{document}